\documentclass[aps,pra,prabib,showpacs,superscriptaddress,twocolumn]{revtex4}
\usepackage{graphicx}
\usepackage{amssymb}
\newcommand{\be}{\begin{equation}}
\newcommand{\ee}{\end{equation}}
\newcommand{\ua}{\uparrow}
\newcommand{\da}{\downarrow}

\newcommand{\eup}{\epsilon_{F \uparrow}}

\newcommand{\Rs}{{R}_{\rm S}}

\begin{document}
\title{The role of interactions in spin-polarised atomic Fermi gases at unitarity}

\author{A. Recati}\email{recati@science.unitn.it}
\affiliation{Dipartimento di
Fisica, Universit\`a di Trento and CNR-INFM BEC Center, I-38050
Povo, Trento, Italy}
\author{C. Lobo}
\affiliation{School of Physics and Astronomy, University of Nottingham, University Park, Nottingham, NG7 2RD, United Kingdom}
\author{S. Stringari}
\affiliation{Dipartimento di
Fisica, Universit\`a di Trento and CNR-INFM BEC Center, I-38050
Povo, Trento, Italy}

\begin{abstract} 
We  study the zero temperature properties of a trapped polarized Fermi gas at unitarity by assuming phase separation between an unpolarized superfluid and a polarized normal phase. The effects of the interaction are accounted for using the formalism of quasi-particles to build up the equation of state of the normal phase with the Monte Carlo results for the relevant parameters. Our predictions for the Chandrasekhar-Clogston limit of critical
polarization and for the density profiles, including the density jump at
the interface, are confirmed with excellent accuracy by the recent
experimental results at MIT. The role of interaction on the radial width of the minority component, on the gap of spectral functions and on the spin oscillations in the normal phase is also discussed. Our analysis points out the Fermi liquid nature of these strongly interacting spin polarized configurations.
\end{abstract}

\maketitle

\section{Introduction}
The behaviour of spin-polarized Fermi systems has been a fascinating subject of research for over fifty years since it can lead to the interplay of superfluidity and magnetism. In particular a Fermi superfluid is resistant to spin polarisation and so one can naturally ask what happens when we attempt to polarize it.

One such example occurs in a superconducting metal when we apply a magnetic field. Under certain conditions the coupling to the orbital motion (responsible for the Meissner effect) is negligible and the important effect is the coupling to the electron spins. The field can lower the energy of the spin-polarized normal state and, if it is strong enough, make the normal state energetically more favourable than the superconducting spin-singlet state. The value of the field at which this transition takes place is known as the Chandrasekhar-Clogston limit \cite{CC} and, in a BCS superconductor, it requires that the field (or, in neutral systems, the chemical potential difference between spin up and down) be larger than $\Delta/\sqrt{2}$ where $\Delta$ is the gap. Crucially, this estimate assumes that the change in energy of the normal state due to polarization is only kinetic in origin and neglects changes in the interaction energy.

However, if the system is strongly interacting, the value of the Chandrasekhar-Clogston field can also depend on the interactions in the normal state and we must accurately take them into account if we wish to study normal/superfluid coexistence.

Exactly this situation has arisen in recent experiments that have been carried out in the strongly interacting unitary limit of two-component atomic Fermi  gases \cite{MITvortices,MIT1,MIT2,Rice1,Rice2}. Such gases can be polarized leading to a state most naturally described as phase separated between a normal and a  superfluid component. The question then revolves around the energetics of such a system. Most of the previous theoretical work has concentrated on the nature of possible superfluid phases and assumed that interactions in the normal state are unimportant \cite{mean-field}. \\\indent
In the present work we analyse in detail the role of interactions in establishing the phase diagram of the unitary spin-polarised system as well as in other relevant phenomena as the density profiles of trapped gases and the frequency shifts of the collective oscillations. We will develop the Fermi liquid theory of the normal state which was introduced in \cite{normal} and show that not only its predictions for the critical polarization (Chandrasekhar-Clogston limit), but also the detailed structure of the density profiles (including the density jump at the interface between the superfluid and normal phases) agree very well with the experimental data obtained at MIT \cite{MIT1,MIT2}. The same theory provides explicit predictions  for the frequencies of the spin excitations in the normal phase. The experimental measurement of these frequencies would provide a direct measurement of the interaction coupling constants characterizing the normal phase itself. \\\indent
We begin in Section \ref{sec:Alessio} by reviewing the theory of the normal state and we study the resulting nature and properties of the superfluid/normal phase transition. Then in Section \ref{sec:Carlos} we review the key points of the MIT experimental situation concentrating on those which are crucial in understanding the phase diagram at very low temperature. Next we make a detailed comparison of theory and experiment, clarifying the features that interactions are responsible for. In Section \ref{sec:Sandro} we study the spin modes in the normal phase and make predictions for their frequency shifts due to interactions in the unitarity limit. Finally in Section \ref{sec:Conclusions} we draw our conclusions.

\section{Interactions in the superfluid and normal phase}
\label{sec:Alessio}

We concentrate on $T=0$ configurations in the unitarity limit, when the scattering length, between the spin-$\ua$ and spin-$\da$ species, is infinite and the system, in many respects, behaves like a strongly interacting fluid. Important features accessible experimentally are related to the polarization of the Fermi gas which permits to explore different quantum phases. In particular recent experiments have revealed the occurrence of phase separation between a unpolarized superfluid and a polarized normal gas \cite{MIT2}.

In what follows we shall consider only two phases: the unpolarized superfluid and the partially polarized normal phase. If not otherwise specified, we will call $n_{\rm S}\equiv n_{{\rm S}\ua}=n_{{\rm S}\da}$ the density in the superfluid phase and simply $n_\ua$ and $n_\da$ the normal state spin-$\ua$ and spin-$\da$ densities.

The equation of state for a homogeneous unpolarized superfluid at unitarity is simply given by 
\begin{equation}
\frac{E_{\rm S}}{N_{\rm S}}= \xi_{\rm S}\frac{3}{5} \frac{\hbar^2}{2m} (6 \pi^2 n_{\rm S})^{2/3}\equiv \epsilon_{\rm S}(n_{\rm S}),
\end{equation}
where $N_{\rm S}$ is the number of atoms in the superfluid phase and the universal parameter $\xi_{\rm S}=0.42$ has  been calculated employing  Quantum Monte Carlo simulations  \cite{Carlsonxi,Stefanoxi}.

A convenient way to build up the equation of state for the normal state is to consider a dilute mixture of spin-$\da$ atoms immersed in a non-interacting gas of spin-$\ua$ atoms.
When the concentration $x=n_\da/n_\ua$ is small the energy of the system
can be written in the form \cite{normal}
\begin{eqnarray}
\frac{E(x)}{N_\ua}&=&\frac{3}{5}\eup\left(1-A
x+\frac{m}{m^*}x^{5/3}+B x^2\right) \nonumber\\
&\equiv&\frac{3}{5}\eup \epsilon(x),  
\label{eq:energyx}
\end{eqnarray}
where $N_\ua$ is the number of spin-$\ua$ atoms and $\eup=\hbar^2/2m (6 \pi^2 n_\ua)^{2/3}$ is the Fermi energy of the spin-$\ua$ gas.
The first term in Eq. (\ref{eq:energyx}) corresponds to the energy per particle of the non-interacting
gas. The linear term in $x$ gives the single-particle energy
of the spin-down particles, while the $x^{5/3}$ term represents the quantum pressure of a Fermi gas of quasi-particles with an effective mass $m^*$. Eventually, the last term includes the effect of interactions between quasi-particles.
The values of the coefficients entering in Eq. (\ref{eq:energyx}) have been the object of various studies using different many-body appoaches\cite{normal,Chevy,Bulgac,sebastiano,Combescot,Prokof'ev}. 
In the present work we use the most recent values $A=0.99(1)$, $m^*=1.09(2)$  and $B=0.14$ calculated in \cite{sebastiano}, using Fixed-Node Monte-Carlo techniques. It is worth remarking that, while Eq. (\ref{eq:energyx}) is thought as an expansion of the normal state energy for small concentration, it agrees very well with Monte-Carlo calculations also for large values of $x$ \cite{normal}.

We assume that the Fermi gas is confined by an isotropic harmonic potential $V({\bf r})=m \omega^2 r^2/2$ 
and that a non-zero polarization \be P=\frac{N_\ua-N_\da}{N_\ua+N_\da} \ee can give rise to a phase separation between a unpolarized superfluid core and an external partially polarized normal shell. We define $\Rs$ the surface which separates the two phases and $R_\ua$ the external surface of the (fully polarized) normal part. In the local density approximation the free energy reads 
\begin{eqnarray}
&E&=2 \int_{r<\Rs} d{\bf r}(\epsilon_{\rm S}(n_{\rm S}({\bf r}))-\mu^0_{\rm S}+ V({\bf r})) n_{\rm S}({\bf r})\nonumber\\
&+&\int_{\Rs<r<R_\ua} d{\bf r}(
3/5\eup\epsilon(x({\bf r}))n_{\ua}({\bf r})\nonumber\\
& &\hspace{0cm}+V({\bf r})(n_\da({\bf r})+n_\ua({\bf r}))-\mu^0_\ua n_\ua({\bf r})-\mu^0_\da n_\da({\bf r}) )
\label{eq:energysplit}
\end{eqnarray}
where $\mu^0_\ua$, $\mu^0_\da$ are the chemical potentials of the spin-$\ua$, spin-$\da$ component respectively and $\mu^0_{\rm S}=(\mu^0_\ua+\mu^0_\da)/2$ is the superfluid chemical potential.

The equilibrium is found by minimizing the energy with respect to the densities of the superfluid and the normal part as well as to the border $\Rs$ between the two phases. By varying the densities we find the LDA expressions for the superfluid and for the normal densities
\begin{equation}
\mu^0_{\rm S}=\xi_{\rm S}\frac{\hbar^2}{2m}(6\pi^2 n_{\rm S})^{{2}/{3}}+V({\bf r})\label{eq:muS},
\end{equation}
\begin{equation}
\mu^0_{\ua}=\left(\epsilon(x)-\frac{3}{5}x\epsilon'(x)\right)
\frac{\hbar^2}{2m}(6\pi^2 n_{\ua})^{{2}/{3}}+V({\bf r}),
\label{eq:locdensmup}
\end{equation}
\begin{equation}
\mu^0_{\da}=\frac{3}{5}\epsilon'(x)\frac{\hbar^2}{2m}(6\pi^2 n_{\ua})^{{2}/{3}}+V({\bf r}).
\label{eq:locdensmud}
\end{equation}
From Eq.(\ref{eq:locdensmup}) we find that the external radius of the normal part is simply given by $R_\ua^2=2 \mu^0_\ua/(m \omega^2)$. 

We are dealing with a polarized system and the same atoms are constituents of both the superfluid and the normal phase, thus the number of particles in the two different phase is not fixed a priori but is determined by the chemical and the mechanical  equilibrium. The latter is obtained by varying the energy functional Eq. (\ref{eq:energysplit}) with respect to $\Rs$, yielding the equal pressure condition between the two phases reads
\begin{equation}
\left[n_{\rm S}^2\frac{\partial{\epsilon_{\rm S}}}{\partial n_{\rm S}}\right]_{r=\Rs}=\frac{1}{2}\left[n_\ua^2\frac{\partial{\epsilon_{\rm N}(x)}}{\partial n_\ua}+n_{\da}^2\frac{\partial{\epsilon_{\rm N}(x)}}{\partial n_\da}\right]_{r=\Rs}.
\label{eq:pressure}
\end{equation}
From Eqs. (\ref{eq:muS}-\ref{eq:pressure}), we obtain an implicit equation for the critical concentration at the border, namely
\begin{eqnarray}
\epsilon(x(\Rs))&+&\frac{3}{5}(1-x(\Rs))\epsilon'(x(\Rs))\nonumber\\
&-&(2\xi_{\rm S})^{3/5}(\epsilon(x(\Rs)))^{2/5}=0.
\label{eq:condtrap}
\end{eqnarray}
Since in the superfluid phase the densities of the spin-$\ua$ and spin-$\da$ component are equal, the occurrence of a solution of Eq. (\ref{eq:condtrap}) with $x(\Rs)\neq 1$ reveals the existence of density jumps at the interface. Using the above-mentioned values for $A$, $m^*$ and $B$ we find $x(\Rs)=0.44$, 
which also corresponds to the critical concentration for a homogeneous system to start nucleating the superfluid phase\cite{normal}.
We find that the jump is significant only in the minority component being
\begin{eqnarray}
\frac{n_{\ua}(\Rs)}{n_{\rm S}(\Rs)}&=&\left(\frac{2\xi_{\rm S}}{\epsilon(x(\Rs))}\right)^{3/5}\sim 1.02,\\
\frac{n_{\da}(\Rs)}{n_{\rm S}(\Rs)}&=&x(\Rs)\left(\frac{2\xi_{\rm S}}{\epsilon(x(\Rs))}\right)^{3/5}\sim 0.45.
\label{eq:jp}
\end{eqnarray}

In addition to the critical concentration Eqs. (\ref{eq:muS})-(\ref{eq:locdensmud}) can be used to calculate the density profiles
which lead us to the important prediction that when the polarization is larger than the critical value $P_c=77\%$ the superfluid core disappears and only the normal phase is present in the trap \cite{normal}. 

A deeper insight on the role of interaction can be obtained in the high polarization case, i.e., $N_\ua\gg N_\da$. Starting from Eq. (\ref{eq:locdensmup}) and (\ref{eq:locdensmud}), to the leading order in $x$, we get 
\be
\mu^0_{\ua}=\frac{\hbar^2}{2m}(6\pi^2 n_{\ua})^{{2}/{3}}+V({\bf r}),
\label{eq:mupHP}
\ee
\be
\mu^0_{\da}+\frac{3}{5}A\mu^0_\ua=\frac{\hbar^2}{2m^*}(6\pi^2 n_{\da})^{{2}/{3}}+V({\bf r})\left(1+\frac{3}{5}A \right),
\ee
yielding the non-interacting value $\mu_\ua^0=(6 N_\ua)^{1/3}\hbar \omega$ for the chemical potential of the majority component.
The above equations reveal that both the majority and the minority components have an ideal Fermi gas profile, the latter being described by a renormalized mass $m^*$ and feeling a renormalized external potential. 
The radius of the minority component is quenched by the interaction to the value
\be
R_\da=R_\da^0 \left[\frac{m^*}{m}\left(1+\frac{3}{5}A\right)\right]^{-1/4},
\label{eq:R_da}
\ee
where $R_\da^0=(48 N_\da)^{1/6}\sqrt{\hbar/(m \omega)}$ is the Thomas-Fermi radius of the ideal Fermi gas. These results can be also understood by introducing the effective single quasi-particle Hamitonian
\be
H_{sp}=-\frac{3}{5}A\mu_\ua^0+ {p^2 \over 2m^*} + V({\bf r}) \left( 1+\frac{3}{5}A \right) .
\label{eq:Hsp}
\ee
for the minority component. 

The above formalism allows to characterize, in terms of the interaction parameter $A$, the energy threshold of the spectral function $\sum_n|<n|c^\dagger_{3,p}c_{\da,p}|0>|^2\delta(\omega+E_n-E_0)$ for the pseudospin flip of the minority component to a third (hyperfine) state $|3>$. This function is relevant for photoemission spectroscopy \cite{JinPhotoem} and RF experiments \cite{MITRFnorm}, where it enters integrated over the momentum $p$ (for recent theoretical calculations for highly polarized samples see \cite{GeorgStoof,SACHDEV}).

In the absence of final state interactions the threshold in the center of the trap is given by 
\be
\Delta({\rm threshold})=\frac{3}{5}A\mu_\ua^0+\frac{p^2}{2 m}\left(1-\frac{m}{m^*}\right),
\label{eq:RFgap}
\ee
which gives in principle a method to measure directly the effective mass.

Other measurable quantities which can give information on the parameters are the frequencies of the spin collective oscillations of the normal phase. Once again in the highly polarized case they can be easily calculated using the Hamiltonian Eq. (\ref{eq:Hsp}) as we discuss in Sec. \ref{sec:Sandro}.

\section{Comparison with experiment}
\label{sec:Carlos}

As the theory developed in the previous section is based on a local density approximation it is important to focus on experiments where such an approach, which ignores details of the interface at the microscopic scale, is applicable. This favours the comparison with the MIT experiments of \cite{MIT1,MIT2} carried out with large values of particle number $N$. Due to strong anisotropy and smaller values of $N$, the Rice experiments reported in \cite{Rice1,Rice2} are instead expected to be much more sensitive to surface effects \cite{ErichSurface} and they will not be considered here. 

The main results carried out at MIT at the lowest temperatures at unitarity can be summarized as follows:

1) There is a critical polarization $P_c\sim 75\%$ above which there is no evidence of superfluidity.
This value has been established by studying vortices \cite{MITvortices} and using the ramp method into the BEC limit to measure the number of condensed molecules. Furthermore the disappearance of the condensate coincides with the disappearance of a central region where the spin-$\ua$ and spin-$\da$ densities are equal \cite{MIT1}.

2) At values of the total polarization $P<P_c$ three regions can be distinguished in the trap: a central superfluid core in which both densities are equal, a first shell where both spin-$\ua$ and spin-$\da$ atoms are present but with different densities, an exterior shell where only spin-$\ua$ atoms are present. 

3)  For $P>P_c$ only two regions are left: an interior partially polarized core and the exterior fully polarized shell.

4) At the interface between the superfluid and the partially polarized phases there is a large jump of around 40\% in the density of the down atoms whereas the discontinuity in the up atoms, if any, is too small to be detected. In going from the partially polarized to the fully polarized phase the densities are continuous. In the spirit of LDA we therefore identify the former transition as first order due to the abrupt density change, whereas the latter one is identified as second order. 

All the four above-mentioned facts agree very well with the theory for the normal phase exposed in the previous section.

We proceed by comparing the theoretical density profiles with the available data\cite{MIT1,MIT2}. 
\vspace{1cm}
\begin{figure}[htbp]
\begin{center}
\includegraphics[width=8cm]{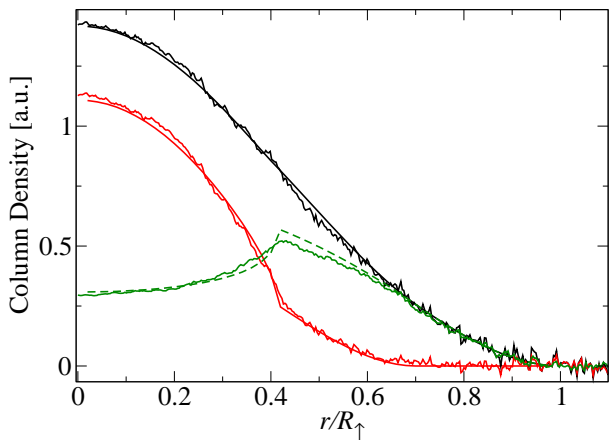}
\caption{Column density profiles for the majority, the minority component and for their difference: MIT data vs. our theory (curves obtained by the method described in the text) for $P=44\%$.}
\label{fig:1}
\end{center}
\end{figure}
In Fig.~(\ref{fig:1}) we compare our theory with the experimental data for the column density for $P=44\%$ and in Fig.~(\ref{fig:2}) we show the comparison with the double integrated density difference  
\begin{equation}
n_d(z)=2 \pi \int_{-\infty}^\infty d\rho \rho (n_\uparrow(z,\rho)-n_\downarrow(z,\rho)),
\ee  
for a range of polarisations both above and below $P_c\sim 77\%$. As it can be observed, the matching is excellent in all cases. 

\begin{figure}[htbp]
\begin{center}
\includegraphics[width=8cm]{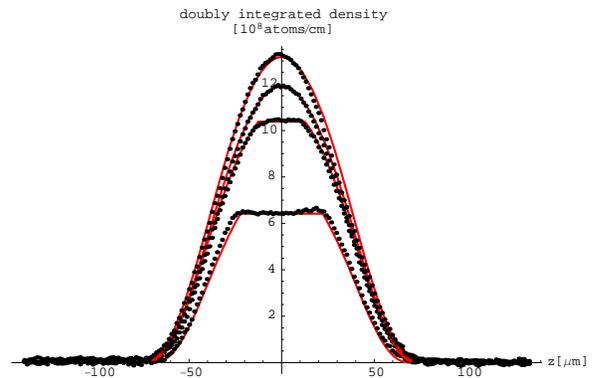}
\caption{Doubly integrated density profile difference ($n_d$), data (black dots) and our theory (red lines) for different polarisations, above and below $P_c$, starting with the lowest curve and moving up: $P=58\%$, $73\%$, $80\%$, $92\%$.}
\label{fig:2}
\end{center}
\end{figure}

The agreement between theory and experiment is also revealed in Fig.~(\ref{fig:compdens}) where we analyse the reconstructed three dimensional density: the jump in the $n_\da$ density is evident (see also \cite{Shin}). 
We emphasize that, since the data are normalized to the non-interacting gas, the only input needed in Figs.~(\ref{fig:1}) and (\ref{fig:compdens}) is the polarization $P$. 
In the case of Fig. (\ref{fig:2}), where the experimental data are not given in dimensionless units, we have used the measured central ($z=0$) density as a fitting parameter. Below the critical polarization the doubly integrated density profiles exhibits a typical plateu whose existence is a direct consequence of the LDA in the presence of an unpolarized superfluid core \cite{Erich}.

We stress that the good agreement is ensured by the proper inclusion of interactions in the normal state. This is not just a quantitative question since assuming that the normal state is noninteracting can lead to unphysically high values of polarisation for the Chandrasekhar-Clogston limit. This is the case of the extensively employed Bogoliubov - de Gennes theory. Although the energy of the superfluid even at unitarity is quite close to the Monte Carlo one, for the normal phase such a theory includes the interactions only through  the pairing terms in the superfluid phase yielding a value close to $100\%$ for the critical polarization\cite{mean-field}, a value definitely ruled out by the MIT experiments. 

\begin{figure}
\begin{center}
\includegraphics[height=5.5cm]{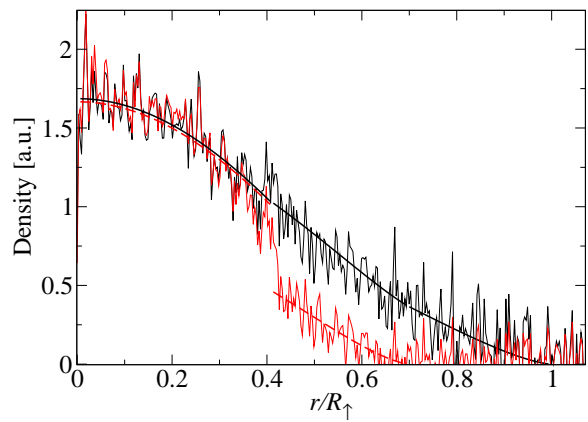}
\end{center}
\caption{Density profiles for a polarization $P=44\%$. Theory: solid black line (dashed red line) is the spin-$\ua$ (spin-$\da$) density. Experiment: the black (red) line is the spin-$\ua$ (spin-$\da$) 
density as reported in \cite{MIT2}. The density jump in the $\da$ component is clearly visible.}
\label{fig:compdens}
\end{figure}

\section{Spin modes}
\label{sec:Sandro}

The excellent agreement with experiments discussed in the previous section  strongly supports the basic picture underlying  this paper, i.e. the assumption that, at unitarity, a phase separation takes place between a superfluid unpolarized core and an external polarized normal phase. It is useful  to further  exploit  the relevant features of this normal phase. In this section we will discuss the  dynamic behavior and we will show, in particular, that the study of the  out of  phase oscillations of the spin-up and spin-down components can provide a further crucial test of the role of interactions in the normal phase.

 A first crucial point to discuss is the role of collisions. Collisions are quenched by Pauli blocking at very low temperature, so that at sufficiently small temperature the small amplitude oscillations in the normal phase are described by collisionless dynamics \cite{decaybruun}. In the opposite collisional regime of relatively higher temperatures the dynamics is instead governed by the equations of hydrodynamics and the out of phase oscillations are overdamped since the spin current is not conserved in this case. Notice that the possibility of studying the collisionless behaviour in the normal phase of a unitary Fermi gas at very low temperature is a unique opportunity provided by spin polarized  Fermi gases. In the non polarized case the gas is in fact always  superfluid at such temperatures.

We will consider the normal phase above the Chandrasekhar-Clogston limit and in particular we will focus on the simplest case of a highly polarized gas where the number of particles in the spin-down component is much smaller than in the spin-up component ($N_\downarrow\ll N_\uparrow $). In this limit  the collective oscillations  can be classified into two cathegories: the "`in phase"' oscillations, where the motion is basically dominated by the majority component, and an "`out of phase"' oscillations where the minority component moves in the trap, the majority one being practically at rest. The most important example of the first class of oscillations is the dipole center of mass motion.  This motion is not affected by the interaction because of the translational invariance of the force and occurs exactly at the frequency $\omega_D= \omega_i$ where $i=x,y,z$. A second important "`in phase"' oscillation is the quadrupole mode. This mode is affected by interaction effects except in the deep $N_\downarrow\ll N_\uparrow $ limit where its frequency approaches  the value $\omega_Q=2\omega_\perp$ (for simplicity we consider only the radial mode in an axi-symmetric trap). Notice that this frequency differs from the hydrodynamic value $\sqrt2\omega_\perp$ \cite{Stringari96}.

Let us now discuss the ``out of phase'' (spin) oscillations. According to the discussion of  Sec. \ref{sec:Alessio} in the $N_\downarrow\ll N_\uparrow $ limit these oscillations are described by the single quasi-particle Hamiltonian Eq.(\ref{eq:Hsp}).

The frequency of the spin oscillations are hence easily calculated. The spin dipole frequency is given by \cite{normal}
\begin{equation}
\omega_D^{(s)}= \omega_i \sqrt{{m\over m^*}\left(1+\frac{3}{5}A\right)}
\label{eq:omD}
\end{equation}
while the spin radial quadrupole frequency is given by
\begin{equation}
\omega_Q^{(s)}= 2\omega_\perp \sqrt{{m\over m^*}\left(1+\frac{3}{5}A\right)}\label{eq:omQ}
\end{equation} 
The above results point out that interactions affect the spin frequencies in a sizable way by increasing the corresponding values in the absence of interaction by a factor $\sim 1.23$.  Their measurement, together with the determination of the radii and/or the spectral function threshold, would consequently provide unique information on the separate value of the relevant interaction parameters $m^*$ and $A$. In this respect it is worth noticing that the quantities $m^*/m$ and $1+(3/5)A$ enter in Eq. (\ref{eq:R_da}) and in Eqs. (\ref{eq:omD}-\ref{eq:omQ}) via a different combination. Notice also that differently from the radius (\ref{eq:R_da}) and the threshold (\ref{eq:RFgap}), the spin frequencies (\ref{eq:omD}) and (\ref{eq:omQ}) are independent on the number of atoms.

The above results can be generalized to include larger values of $N_\downarrow/N_\uparrow$, through the proper inclusion of the interaction term between the minority and majority species in the energy functional introduced in Sec.~\ref{sec:Alessio}.

An important question is how these spin modes can be excited experimentally. It is worth noticing that an external perturbation coupled to the total density $n=n_\uparrow + n_\downarrow$ will never excite the spin dipole mode, but only the center of mass motion which is a pure density oscillation, insensitive to two-body forces (Kohn's theorem). Since at the high magnetic fields characterizing the resonant regime the magnetic coupling is practically the same for the two hyperfine states of $^6$Li used in the experiments, only an optical coupling with a laser field suitable detuned with respect to the internal atomic frequencies of the two hyperfine states can consequently excite the dipole spin mode. The situation is different for the quadrupole oscillation. In fact in this case, due to the absence of the analog of the Kohn's theorem, the spin excitation is not completely decoupled from the density probe and a quadrupole perturbation of the form $\sum_i x_iy_i$  will result in the excitation not only of the ``in phase'' mode, but also of the spin mode (\ref{eq:omQ}) with a relative weight proportional to $<r^2_\perp>_\downarrow N_\downarrow / <r^2_\perp>_\uparrow N_\uparrow$. 

\section{Conclusions}
\label{sec:Conclusions}

By assuming phase separation between a superfluid and a polarized normal gas, we have provided an accurate picture of the zero temperature behavior of a trapped polarized Fermi gas at unitarity. Using the recent results of {\it ab initio} Monte Carlo calculations for  the relevant interaction parameters, in both the superfluid and the normal phase, we have found excellent agreement with experiments both concerning the value of critical polarization (Chandrasekhar-Clogston limit) and the shape of the density profiles,  including the density discontinuity at the interface between the two phases.  The agreement confirms that the basic structure of these novel configurations is now well understood. In particular it shows that the proper inclusion of the interactions is a crucial requirement for a quantitatively and qualitatively correct description of the polarized phase which emerges like a new  example of strongly interacting  normal Fermi liquid, consisiting of a weakly interacting gas of quasi-particles and remaining normal even at the lowest temperatures.

The excitation of the  spin oscillations in the collisionless regime (see discussion in Sec. \ref{sec:Sandro}), the measurement of the radial width of the minority component (see Eq. (\ref{eq:R_da})), the availability of photoemission spectroscopy \cite{JinPhotoem} and RF measurements \cite{MITRFnorm} and the behaviour   of the  gas under the constraint of an adiabatic rotation \cite{Ingrid} are expected to provide further insight on the role of the interactions  in these polarized Fermi gases. 

We finally point out that there are a number of issues related to the border between the superfluid and the normal phase. In particular when mesoscopic effects become important (small number of atoms, strong trap deformation) the proper description of the interface, beyond LDA, can play an important role in characterizing the phase separation between the superfluid and the normal component.

\section*{Acknowledgemt}
We gratefully thank Yong Shin for kindly providing us with the experimental data and Ingrid Bausmerth for carefully reading the paper.

\end{document}